\documentclass[aps,prc,twocolumn,showpacs,preprintnumbers,amsmath,amssymb,floatfix]
              {revtex4}

\usepackage{graphicx}           

\begin{document}
 
\preprint{\fbox{\sc version of March 30, 2003}}
 
\def\lt{\raisebox{0.2ex}{$<$}}
\def\gt{\raisebox{0.2ex}{$>$}}
 
\title{Mean-Field Description of Fusion Barriers with Skyrme's Interaction}

\author{    A. Dobrowolski$^1$, K. Pomorski$^{1,2}$, J. Bartel$^2$
\\
{\it $^1$Katedra Fizyki Teoretycznej, Uniwersytet M. C. Sk\l odowskiej,
                               Lublin, Poland}
\\
      {\it $^2$IReS -- IN$_2$P$_3$ -- CNRS and Universit\'e Louis Pasteur,
                             Strasbourg, France}}                               


\begin{abstract}

Fusion barriers are determined in the framework of the Skyrme energy-density 
functional together with the semi-classical approach known as the 
Extended Thomas-Fermi method. The barriers obtained in this way with the 
Skyrme interaction SkM$^*$ turn out to be close to those generated 
by phenomenological models like those using the proximity potentials.
It is also shown that the location and the structure of the fusion barrier 
in the vicinity of its maximum and beyond can be quite accurately described 
by a simple analytical form depending only on the masses and the relative 
isospin of target and projectile nucleus. 
\end{abstract}  
\pacs{PACS numbers: 21.60.Jz,21.10.Dr,21.10.-k,21.10.Pc} 

\maketitle

\section{Introduction} 

The knowledge of the collective potentials between two colliding ions is 
absolutely crucial for the synthesis of new isotopes. This problem has been the 
subject of a very active research over the last decade and remains one of the 
most intensively studied subjects in low-energy nuclear physics in particular in
the perspective of the synthesis of super-heavy elements as well as of exotic 
nuclei far away from the $\beta$-stability line. It has in particular been 
shown that models based on a macroscopic approach such as the Liquid Drop Model 
(LDM) or of semi-classical type like the Extended Thomas-Fermi (ETF) method 
together with the Skyrme energy-density functional are able to reproduces quite 
accurately experimental data on fission and fusion barriers.

The research concerning the interaction potential between colliding ions 
goes back to the work of R.\ Bass (for a review see ref.\ \cite{Bass}).
Since the idea of a proximity potential due to W.\ J.\ \'Swi\c{a}tecki and coworkers 
in the late 1970's \cite{Blocki77}, many improvements have been proposed to 
make this phenomenological approach more realistic and general, in particular 
by taking into account the local curvature of the surfaces of target and 
projectile \cite{Frobrich98,Gonchar2002}. One of the main challenges was, as 
already mentioned, to give a reliable guideline for the formation and stability 
of super-heavy elements.

The parameters of the proximity function are usually fitted to experimentally 
known fusion barriers heights. 
The basic idea of all proximity models is to determine the potential between 
the two colliding nuclei as function of the minimal distance $s$ of their 
surfaces and their so-called reduced radius defined in the case of 
spherical nuclei as $\bar{R} = R_{1} \, R_{2} / (R_{1} \!+\! R_{2})$, where 
the indices 1 and 2 refer to the target and projectile nucleus respectively.

Experimental data confirm the existence of a pocket in the shape of the 
nuclear part of the potential. This feature plays an essential role for 
the formation probability but also for the stability of the compound nucleus 
in the fusion experiments. Indeed, if the pocket is deep and wide, several 
quasi-bound states might exist and the probability of forming that compound 
nucleus is large. If, on the other hand, this minimum is shallow and narrow 
no such states will exist. It would therefore be of great help to have at 
hand a simple, yet sufficiently accurate phenomenological expression for the 
determination of fusion barriers in order to evaluate the most favorable 
reaction for the formation of a given compound nucleus.

The aim of our present investigations is to show that the shape of the 
proximity potential can be, indeed, quite accurately determined through 
self-consistent semi-classical calculations. We proceed in a way similar to 
that suggested in Ref.\ \cite{PD80} and recently used in Ref.\ \cite{DN02}. 
After introducing the Skyrme Hartree-Fock energy-density functional together 
with the semi-classical ETF approximation in Section II we explain in Section 
III how this approach can be used to calculate fusion barriers. In Section IV 
the fusion barriers obtained in this way for a sample of 269 different 
reactions (different combinations of target and projectile) leading a priori 
to different isotopes of the super-heavy elements with even charge 
numbers in the range $Z \!=\! 108 \!-\! 114$ are presented and compared to the 
barriers obtained in other theoretical models. Based on these fusion-barrier 
calculations we present in Section V a simple analytical expression for this 
potential and show that it is able to reproduce very accurately the fusion 
barrier heights obtained in the ETF approach. Conclusions are finally drawn 
in Section VI. 

%
%
\section{The Skyrme ETF approach to nuclear structure}  

For the calculation of the interaction potential between the two nuclei we 
need an energy-density functional which has proven its capacity to correctly 
describe nuclear ground-state properties. As we are interested in the potential 
energy resulting from the interaction of the tails of the density distributions 
of target and projectile, a very precise description of the nuclear surface 
seems absolutely crucial. The Skyrme effective interaction has been shown to 
have this property \cite{BFN75,BQ82,CB98} and has been, together with Gogny's 
force \cite{DG80}, the most successful effective nucleon-nucleon interaction  
over the last three decades. Its energy density is given by
\begin{eqnarray} 
E&&= \!\int\! d^3 r \; {\cal E} \left[ \rho_q, \tau_q, \vec{J}_q \right]
    = \!\int\! d^3 r \left\{ \frac{\hbar^2}{2m} (\tau_n + \tau_p)  
                                                          \right.    \nonumber\\
&&+ B_1 \rho^2 + B_2(\rho_n^2+\rho_p^2) + B_3 \rho \tau 
   + B_4(\rho_n \tau_n + \rho_p \tau_p)
                                                                     \nonumber\\
&&- B_5(\vec\nabla \rho)^2 - B_6[(\vec\nabla \rho_n)^2 
   + (\vec\nabla \rho_p)^2]                                      \label{edens}\\
&&+ \rho^{\alpha}[B_7 \rho^2 + B_8(\rho_n^2 + \rho_p^2)]             \nonumber\\
&&- B_9[\vec J \cdot \vec\nabla \rho + \vec J_n \cdot \vec\nabla \rho_n 
    + \vec J_p \cdot \vec\nabla \rho_p] + {\cal E}_{Coul}(\vec{r}\,)  
                       \left.\frac{ }{ } \right\}\,\,,                 \nonumber 
\end{eqnarray} 
where the constants $B_i$ are combinations of  the usual Skyrme-force 
parameters $t_j$ and $x_j$. $\frac{\hbar^2}{2m} \, \tau_q$ and $\vec{J}_q$ 
designate the kinetic 
energy density and spin-orbit density of charge state $q, \; q=\{n,p\}$ 
and non indexed quantities correspond to the sum of neutron and proton 
densities such as $\rho = \rho_n + \rho_p$.

Since our aim is to evaluate fusion barriers, a quantity which is essentially 
determined by the radii and tails of the density distributions of target and 
projectile, we have chosen the semi-classical approximation known as the 
Extended Thomas-Fermi (ETF) method \cite{BG85} to determine in a 
self-consistent way the structure of projectile and target nuclei.

The ETF approach allows to express $\tau_q$ and $\vec{J}_q$ as functions of 
the local density $\rho_q$ and its derivatives. This functional expression has 
the form of an expansion with Planck's action quantum $\hbar$ as order 
parameter. For problems which don't break time-reversal symmetry (as would 
e.g.\ be the case for rotating nuclei) only even powers of $\hbar$ appear 
so that the semi-classical expansion of the kinetic energy density will  
take the form
\begin{equation} 
   \tau^{}_q[\rho^{}_q] = \tau_q^{(TF)}[\rho^{}_q] + \tau_q^{(2)}[\rho^{}_q] 
                        + \tau_q^{(4)}[\rho^{}_q] 
                                   \; ; \;\; q = \{ n,p \} \;.
\label{taurho}\end{equation} 
At lowest order in the ETF expansion one has the well-known Thomas-Fermi 
kinetic energy density 
\begin{eqnarray}
  \tau_q^{(TF)}[\rho_q] = \frac{3}{5}(3 \pi^2)^{2/3} \; \rho_q^{5/3} \;\; ,
\label{tautf}\end{eqnarray} 
whereas the second-order contribution is given by~:
\begin{eqnarray}
   \tau_q^{(2)}[\rho_q]&&= \frac{1}{36} \frac{(\vec{\nabla}\rho_q)^2}{\rho_q}
     +\frac{1}{3} \triangle\rho_q+ \frac{1}{6} \frac{\vec{\nabla} \rho_q \cdot 
       \vec{\nabla} f_q}{f_q}                                   \nonumber \\
&&+ \frac{1}{6} \rho_q \frac{\triangle f_q}{f_q} 
     - \frac{1}{12} \rho_q \left( \frac{\vec{\nabla} f_q}{f_q} \right)^2   \\
&&+ \, \frac{1}{2} \left( \frac{2m}{\hbar^2} \right)^2 
			\rho_q \left( \frac{\vec{W}_q}{f_q} \right)^2 \;\; .
\nonumber
\end{eqnarray} 
Here the effective-mass form factor $f_q \!=\! m / m^*(\vec{r}\,)$, is given 
by 
\begin{eqnarray}
  f_q(\vec{r}\,)&&=\, \frac{m}{m_q^\star(\vec{r}\,)}
    \,=\, \frac{2m}{\hbar^2}\left(\frac{\delta {\cal E}(\vec{r}\,)}{\delta 
                                                 \tau_q(\vec{r}\,)}\right)\\
&&=\, 1 + \frac{2m}{\hbar^2} \, [B_3\rho(\vec{r}\,) + B_4 \rho_q(\vec{r}\,)]
\nonumber
\label{effmass}
\end{eqnarray} 
and the spin-orbit potential as
\begin{eqnarray}
   \vec W_q(\vec{r}\,) 
      \,=\, \frac{\delta {\cal E}(\vec{r}\,)}{\delta \vec J_q(\vec{r}\,)}
      \,=\, -B_9 \, \vec\nabla(\rho+\rho_q) \;\;.
\end{eqnarray}
In the ETF approach one also obtains a semi-classical expansion for the 
spin-orbit density. Since the spin is a pure quantal phenomenon which does 
not have any classical analog, the semi-classical expansion of 
$\vec{J}$ starts at order $\hbar^2$ and the leading term is given by 
\begin{eqnarray}
  \vec J_q^{(2)}=-\frac{2m}{\hbar^2} \, \frac{\rho_q}{f_q} \, \vec W_q \;\;. 
\label{jrho}\end{eqnarray}
This now, however, means that one is able to express the total energy 
density ${\cal E}$ as a unique functional of the proton and neutron density 
alone and to perform a variational calculation of the ground-state properties 
of a nucleus, where the variational quantities are $\rho_n(\vec{r}\,)$ and 
$\rho_n(\vec{r}\,)$. That such a unique functional must exist for a system 
of interacting Fermions has been shown by Hohenberg and Kohn \cite{HK64}. 
In the general case this functional is, however, perfectly unknown. It is 
the achievement of the ETF approach to derive this functional in the 
semi-classical limit in a systematic way. ETF variational calculations have 
indeed proven, when the semi-classical expansion is carried through to order 
$\hbar_{}^4$ \cite{BG85,BB02}, extremely successful to describe average 
(liquid-drop type) nuclear properties (see Ref.\ \cite{BG85} for a review). 
It has in particular been shown \cite{BG85,BB85,CPV90} that modified Fermi 
functions of the form 

\begin{eqnarray}
  \rho_q(r) = \rho_0^{(q)} \, \left( 1 + e^{\frac{r - R_q}{a_q}} 
                \right)^{\!\!-\gamma_q} 
\label{densy}\end{eqnarray}
are in that case a very good approximation to the full variational solution. 
A systematic analysis of the variation of the density parameters 
$\rho_0^{(q)}$, $a^{}_q$ and $\gamma^{}_q$ (the parameter $R^{}_q$ is then 
determined through particle-number conservation) which are the variational 
parameters of the ETF approach, with the mass number $A$ and the isospin 
parameter $I$ can be found in Ref.\ \cite{BPB02}.

We have carried out these kind of 4$^{\rm th}$ order ETF calculations, to 
determine the structure of target and projectile nucleus. The densities 
obtained in this way are then used to determine the fusion potential between 
the two nuclei as will be explained in the following.
%
%
\section{The ETF fusion potential}

For evaluating the fusion potential we will use the {\it sudden approximation}, 
i.e.\ keeping the densities of the colliding ions fixed and neglecting all 
possible rearrangement effects. At first sight, this looks like a quite crude 
approximation. For a fusion reaction the beam energy per nucleon is, indeed, 
quite small as compared to the Fermi energy and an adiabatic treatment would 
be more appropriate. One has, however, to keep in mind that the essential 
characteristics of the fusion barrier, as its structure around the above 
mentioned {\it pocket}, are determined at distances larger and close to
the touching configuration. Defining the minimal distance $s$ between 
the equivalent sharp surfaces of the liquid drops identifying the two nuclei 
we see that the essential features of the fusion barrier or cross-sections 
are determined at the distances $(s > 0)$, where the sudden approximation 
seems to be a very reasonable approach. We can then evaluate the nuclear part 
of the interaction  potential $V_{nuc}$ of two spherical colliding ions as 
function of the distance $d$ between theirs centers of mass (see Fig.\ 1) as~: 
\begin{figure}[h]
  \begin{center}
  \includegraphics[width=7.5cm, angle=00]{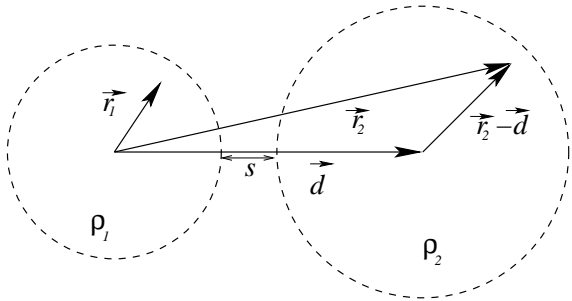}
  \end{center}
  \caption{Definition of target and projectile coordinates.}
  \label{fig1.hip}
\end{figure}
\begin{eqnarray}
   V_{nuc}(d)&&= \int \left\{ {\cal E}^{}_{nuc}[\rho^{(1)}(\vec{r}\,)
      +\rho^{(2)}(\vec{r}- \vec{d}\,)] \right.                           \\
    &&\left. - {\cal E}_{nuc}[\rho^{(1)}(\vec{r}\,)] 
      - {\cal E}_{nuc}[\rho^{(2)}(\vec{r} - \vec{d}\,) ]
       \right\} \, d^3r \;\;,
\nonumber
\label{eq-3.1}\end{eqnarray} 
where ${\cal E}_{nuc}$ is the nuclear part of the energy-density, Eq.\ 
(\ref{edens}),  
and $\rho^{(i)}(\vec{r}\,)$ denotes the density distribution of nucleus $i$ 
alone ($(1)$ labeling  e.g.\ the target and $(2)$ the projectile).

If the distance $d$ between the two nuclei is much larger than the sum of 
the half density radii ($d \gg R^{}_1 + R^{}_2$) then the nuclear part of the 
interaction should vanish because of the short range of the nuclear forces:
\begin{eqnarray}
  \lim_{d \rightarrow \infty}&&{\cal E}^{}_{nuc} \left[ 
                \rho_{}^{(1)}(\vec{r}\,) + \rho_{}^{(2)}(\vec{r} - \vec{d}\,) 
                                 \frac{}{}\! \right] =    \\
 &&=\, {\cal E}^{}_{nuc} \left[ \rho_{}^{(1)}(\vec{r}\,) \frac{}{} \! \right]
      +{\cal E}^{}_{nuc} \left[ \rho_{}^{(2)}(\vec{r}\,) \frac{}{} \! \right]\;.
\nonumber
\label{eq-3.2}\end{eqnarray} 
Thus, the nuclear part of the interaction potential (\ref{eq-3.1}) between 
the nuclei is non negligible only when the distance $s$ between the nuclear 
surfaces of both nuclei is smaller then a few fm. 
The Coulomb interaction between the two ions is given as usual in the form
\begin{equation} 
   V_{Coul}(d) = \int \frac{\rho_{ch}^{(1)}(\vec{r}^{}_1) \; 
         \rho_{ch}^{(2)}(\vec{r}^{}_2)}{|\,\vec{r}^{}_1 - \vec{r}^{}_2 \,|} 
                                                 \; d^3r^{}_1 \, d^3r^{}_2 
\label{eq-3.3}\end{equation} 
where we simply assume that the charge density is given in terms of the 
proton density by $\rho_{ch}^{(i)} \approx \,e \, \rho_{p}^{(i)}$ where the 
latter can be approximated in the ETF approach by Eq.\ (\ref{densy}).
%
%
\section{Fusion barriers for super-heavy elements}

The fusion barrier appears as the result of the competition between  the 
long-range repulsive Coulomb interaction $V_{Coul}$ and the short-range 
attractive nuclear forces $V_{nuc}$ as is illustrated on Fig.~2, where the 
total fusion barrier is plotted as function of the distance $d$ (see Fig.\ 1) 
of the centers of mass of projectile and target nucleus as well as of the 
distance $s$ between their half-density surfaces. One notices that for 
distances $s \geq 3$ fm the attractive nuclear part of the fusion potential 
almost vanishes and the barrier is just determined by the repulsive Coulomb 
potential. Taking the diffuseness of the charge-density distribution into 
account reduces the height of the fusion barrier only by about 1 MeV as 
compared to the barrier obtained with point charge distributions of target 
and projectile. 
\begin{figure}[h]
  \begin{center}
  \includegraphics[height=8.5cm, angle=-90]{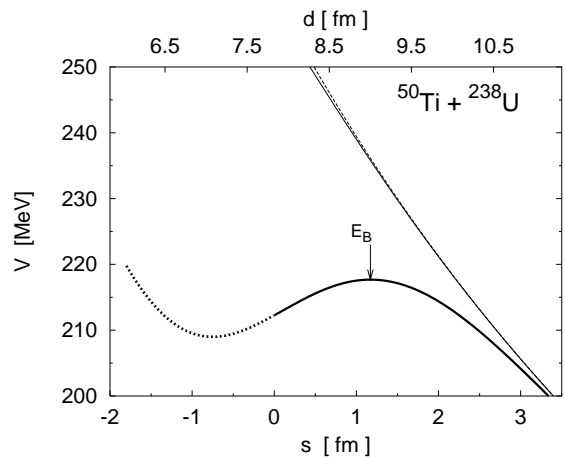}
  \end{center}
  \caption{Shapes of total (nuclear + Coulomb) fusion barriers $V$ (full line)
       as function of the center-of-mass distance $d$ (top) and the distance
       between the equivalent sharp surfaces $s$ (bottom) for the reaction
       $^{50}$Ti + $^{238}$U leading to the super-heavy element $Z \!=\! 114$.
       Also shown are the exact Coulomb barrier (thin line) and the one
       corresponding to two point charges (dashed line).}
  \label{fig2.hip}
\end{figure}
\begin{figure*}[ht]
  \begin{center}
  \includegraphics[height=17cm, angle=-90]{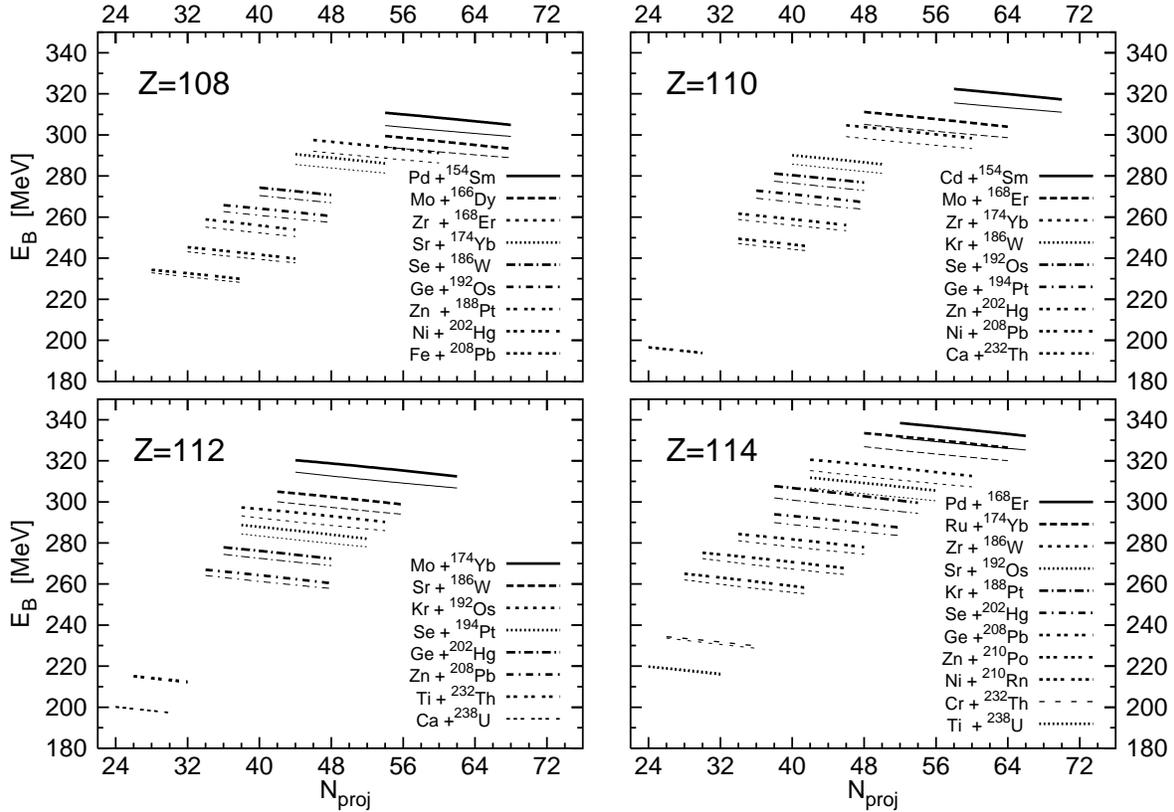}
  \end{center}
  \caption{Height of fusion barriers corresponding to the different reactions
       leading to different isotopes of the S.H.\ elements with
       $Z \!=\! 108,110,112,114$, evaluated in our ETF approach (thick line)
       and in the proximity model of Ref.\ \protect\cite{MS00} (thin lines),
       as function of the projectile neutron number.}
  \label{fig3.hip}
\end{figure*}

It is evident that when the overlap of the densities of the colliding nuclei 
becomes large, as e.g. when $s \leq 0$ (dotted line in Fig.\ 2) the sudden 
approximation used here becomes more and more questionable. Our aim, however, 
is not to give a precise description of the entire fusion potential, i.e.\ 
also for large negative $s$ values, but rather to obtain some reasonable 
estimates of the height $E_B$ of the fusion barrier.
Fig.~3 shows $E_B$ obtained in the sudden approximation and neglecting 
deformation effects of the colliding ions for 269 different reactions (with 
$\beta$-stable targets ranging from $^{168}$Er to $^{238}$U) leading to 
different isotopes of the super-heavy elements with even charge numbers 
between $Z \!=\! 108$ and $Z \!=\! 114$. The barriers shown are the ones 
obtained with the Skyrme forces SkM$^*$ \cite{BQ82} (barriers corresponding 
to the Skyrme SLy4 interaction \cite{CB98} are identical to within 1 MeV). 
The thin lines illustrate the heights of the fusion barriers obtained in the
Myers-\'Swi\c{a}tecki proximity model \cite{MS00}. Barrier heights evaluated using 
the prescription of Ref. \cite{MR01} are very close to the former differing 
by not more than 2~MeV for all the reactions considered here.

%
\section{A simple analytical fusion potential}

The nuclear part of the potential between two colliding ions can be quite well 
reproduced by the approximate form
\begin{equation}
  V^{}_{nuc}(d\,) \approx \tilde{V}^{}_{nuc}(d\,) 
                             = V^{}_0 \;\, e^{-\alpha \; (d - d^{}_0)^2 } \;,
\label{eq-5.1}\end{equation}
where $d$ is the distance between the centers of mass of target and 
projectile and $d_0$ the distance for which the nuclear potential has its
minimum. $d_0$ and the depth $V_0$ of the potential are a direct result of 
our semi-classical calculation. These two quantities are parametrized in the 
following way as functions of the masses $A_1$, $A_2$ and reduced isospins 
$I_i = (A_i - 2 Z_i)/A_i\;$ of target and projectile nuclei 
\begin{equation}
   d_0(A_1,A_2) 
     = r^{}_0 (A_1^{1/3} \!+ \!A_2^{1/3}) + b \;\; ,
\label{eq-5.2}\end{equation}
and 
\begin{equation}
  V_0(A_1,A_2,I_1,I_2) = v^{}_0 \left[1 + \kappa\,(I_1+I_2) \frac{}{}\! \right]
          \frac{A_1^{1/3}\! \,\cdot\, \!A_2^{1/3}}{A_1^{1/3} + A_2^{1/3}} \;\;,
\label{eq-5.3}\end{equation}
where the values of the parameters entering into these expressions can be 
adjusted to the ETF nuclear potentials.

To obtain the potential that the system of two colliding nuclei experiences 
we have to add the Coulomb potential, Eq.\ (\ref{eq-3.3}), to the nuclear 
part. For distances large compared to the sum of the equivalent-sharp-surface 
radii, this Coulomb potential is very well approximated by the Coulomb 
potential between two point charges, but as soon as the densities of the two 
nuclei interpenetrate this approximation will no longer give good results and 
it turns out that the difference 
\begin{equation}
  \Delta V^{}_{Coul}(d\,) \,=\, V^{}_{Coul}(d\,) 
                                      - \frac{Z^{}_1 \, Z^{}_2 \, e_{}^2}{d}
\label{eq-5.4}\end{equation} 
will be, in the case of the super-heavy elements considered here, of the order 
of 3 MeV already in the vicinity of the touching configuration ($s = 0$).

Instead of calculating the diffuse-surface Coulomb energy explicitly for each 
reaction under consideration, the idea is to develop an easy-to-use expression 
to approximate in a reliable way the ETF fusion potential between two nuclei 
developed above, including the full Coulomb potential. That is why we propose 
to write this potential in the form
\begin{equation}
   \tilde{V}(d\,) \,=\, 
           \tilde{V}^{}_{nuc}(d\,) + \frac{Z^{}_1 \, Z^{}_2 \, e_{}^2}{d} \;\;,
\label{eq-5.5}\end{equation} 
where the term $\tilde{V}^{}_{nuc}$, which now contains the Coulomb-energy 
difference $\Delta V^{}_{Coul}$, Eq.\ (\ref{eq-5.4}), in addition to the 
nuclear potential (\ref{eq-3.1}), is approximated by Eq.\ (\ref{eq-5.1}) with 
the parameters $d_0$ and $V_0$ being given by Eqs. (\ref{eq-5.2}) and 
(\ref{eq-5.3}). 
It is clear in this context that these {\it new} parameters can turn out to 
take values that could be quite different as compared to those describing 
the nuclear part alone.

To determine these quantities we perform a simultaneous fit of the 5 free 
parameters to the fusion potentials of the 269 reactions obtained in the 
above described ETF approach (see Fig.\ 3). We proceed in the following way:
What we are interested in is primarily the precise description of the location 
$d^{}_{max}$ and the height $V^{}_{max}$ of the total fusion barrier, and we 
therefore attach a maximum weight to their best possible reproduction. When 
going to smaller distances, i.e.\ towards negative values of $s$, the 
Coulomb-energy difference $\Delta V^{}_{Coul}$, Eq.\ (\ref{eq-5.4}), grows 
larger, but also the nuclear potential obtained in our sudden approximation 
becomes less and less reliable. That is why we attach a decreasing importance 
to the reproduction of the fusion barriers for smaller and smaller distances. 
For increasing distances $d \!>\! d^{}_{max}$, on the other hand, both the 
nuclear potential $V^{}_{nuc}$ as well as the Coulomb correction 
$\Delta V^{}_{Coul}$ go rapidly to zero and our appro\-xi\-mation form 
(\ref{eq-5.5}) becomes better and better. We therefore attach again less 
weight to the precise reproduction of the values with large positive values 
of $s$. We have thus imagined a least-square-fit procedure with a normalized 
weight function of Gaussian form and with a width of 1 fm centered at 
$d \!=\! d^{}_{max}$. The values of the parameters obtained in this way are 
listed in Table 1.

\begin{center}
\begin{tabular}{||c|c|c|c|c||}
\hline
 $\;r^{}_0$ [fm]$\;$ & $\;b$ [fm]$\;$ & $\;v^{}_0$ [MeV]$\;$ 
                     & $\kappa$ & $\;\alpha$ [fm$^{-2}$]$\;$ \\
\hline
 1.183 & -2.400 & -46.07 & $\;$-0.4734$\;$ & 0.173 \\[  1.5ex]
\hline
\end{tabular}
\vspace{3mm}
\noindent

Tab.\ 1: Values of parameters entering the fit of the fusion potential, 
Eq.\ (\ref{eq-5.5})
\end{center}

The final criterion of the accuracy of our approach consists in its ability 
to reproduce the height and shape of the fusion barriers for all 269 reactions 
obtained within the ETF approach with the SkM$^*$ Skyrme interaction. 
The r.m.s. deviation of the exact ETF fusion potential and its approximation 
by Eq.\ (\ref{eq-5.5}) is only of 0.27 MeV at the top of the barrier and of 
0.37 MeV for the touching configuration ($s = 0$).

Such a fusion barrier is shown on Fig.\ 4 for the reactions 
$^{48}$Ca$ + ^{232}$Th. 
As one can see, the ETF barrier is almost perfectly reproduced by our 
analytical expression, Eq.\ (\ref{eq-5.5}) with (\ref{eq-5.1}). The 
barrier height is also almost the same as that obtained in the proximity 
model of Ref.\ \cite{MS00}.
\begin{figure}[h]
  \begin{center}
  \includegraphics[width=8.5cm, angle=00]{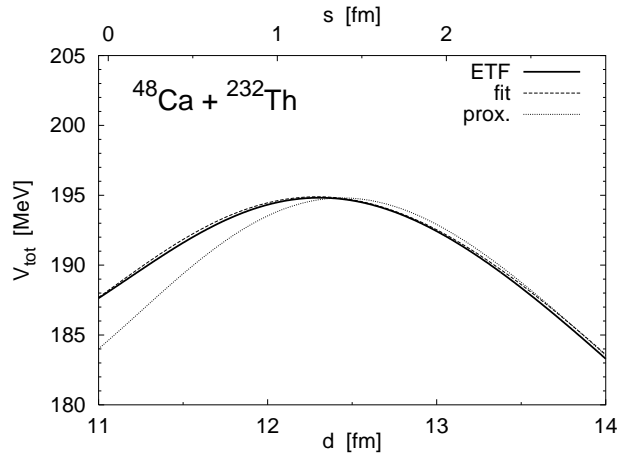}
  \end{center}
  \caption{Fusion barrier for the reaction $^{48}$Ca + $^{232}$Th obtained within
       the ETF approach (full line), its approximate analytical form, Eq.\
       (\protect\ref{eq-5.5}) (dashed line) and the proximity potential of
       Ref.\ \protect\cite{MS00} (dotted line).}
  \label{fig4.hip}
\end{figure}

It is obvious from Fig.\ 3 that the agreement between our ETF approach and 
the proximity model is not always going to be that close. If one considers 
e.g. the reaction $^{110}$Cd + $^{154}$Sm one already concludes from Fig.\ 3 
that the barrier heights of the two approaches are going to be different 
by some 7 MeV (or 2\%). One also notices that the minimum obtained in our 
ETF approach is quite shallow, whereas the proximity potential does not 
make any predictions about the region $s < 0$. One could now speculate 
about the validity of our approach close to and inside the touching 
configuration. The enhancement of the height of the fusion barriers related 
to the sudden approximation is obviously related to the concept of the 
{\it diabatic fusion barriers} advocated by W.\ N\"orenberg and co-workers 
\cite{BLN89}.
\begin{figure}[h]
  \begin{center}
  \includegraphics[width=8.5cm, angle=00]{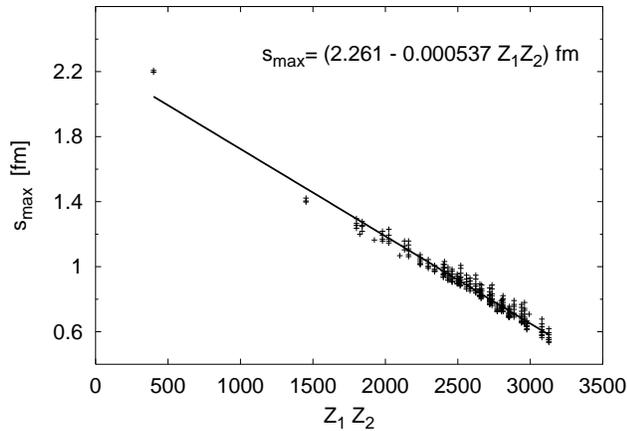}
  \end{center}
  \caption{Location $s^{}_{max}$ of the top of the fusion barrier as function
       of the product $Z^{}_1 \, Z^{}_2$ of the charge numbers of target and
       projectile.}
  \label{fig5.hip}
\end{figure}

The coincidence of the ETF fusion barrier heights with those obtained in the 
proximity approach for very asymmetric reactions ($A_1 \gg A_2$) seems to be 
related to the value $s^{}_{max}$ of the distance $s$ of the equivalent sharp 
surfaces of the two colliding nuclei at the top of fusion barrier, as this is 
demonstrated on Fig.\ 5. Comparing Figs.\ 3 and 5 one can, indeed, observe 
that the enhancement of the ETF barrier height relative to prediction of the 
proxi\-mity model is larger as the distance of the sharp surfaces of the two 
nuclei at the top of the barrier gets smaller which is the case for the 
reactions between nearly symmetric nuclear systems.

As we have just considered fusion reactions leading to the super-heavy 
elements with even values of $Z$ between $Z \!=\! 108 - 114$ one might doubt 
about the utility of such an analysis for the description of fusion barriers 
in other regions of the atomic chart. We have therefore applied our 
phenomenological expression to the reactions 
$^{160,162,166}$Dy + $^{46-50}$Ti as well as to the very light system 
$^{48}$Ca + $^{48}$Ca. To our great surprise the predictions based uniquely 
on our knowledge of the ETF barriers of the super-heavy system gave 
astonishingly good results for these lighter systems with a deviation between 
ETF results and those of the simple analytical expression, Eq.\ (\ref{eq-5.5}),
of less than 0.7 MeV for the former and of about 1.3 MeV for the latter system.

%
%
\section{Conclusions}

We have presented a model allowing for a systematic investigation of fusion 
barriers between heavy nuclei. It has been demonstrated that our approach 
which is based on a self-consistent semi-classical description of projectile 
and target nucleus and has {\it no adjustable parameter} predicts fusion 
barriers that are quite close to the ones obtained with the phenomenological 
proximity approach of Myers and \'Swi\c{a}tecki \cite{MS00}. We have given, in 
addition, an analytical form allowing for a simple approximation of the ETF 
fusion barriers, thus providing a very simple evaluation of fusion potentials 
for reactions throughout the periodic table. We believe that such an approach 
could serve as a guide-line for the research of the synthesis of super-heavy 
elements.

It is also quite remarkable that our simple phenomenological ansatz, Eq. 
(\ref{eq-5.5}), adjusted to the reactions leading to super-heavy elements 
was able to reproduce even the fusion barrier of very light systems in a quite 
satisfactory way with a relative error of 2$\,\% 
$ at most. 

One might also speculate about the importance of deformation on the barrier 
heights obtained in our approach. It seems immediately evident that taking 
into account deformation as an additional degree of freedom of the system 
target-projectile could often lead to a decrease of the fusion barrier 
\cite{PPR94}. One knows, indeed, that e.g. the Coulomb barrier is 
substantially lower for two nuclei with ellipsoidal deformation and tip-to-tip 
orientation as compared to a different orientation or in the absence of 
deformation. It is obvious that for a system like $^{110}$Cd + $^{154}$Sm, 
which we briefly discussed above, where at least the target nucleus shows a 
substantial deformation, such an effect plays a non negligible role which 
implies that taking deformation into account in our ETF approach might lead 
to much lower fusion barriers for those systems. 

\noindent \\
{\bf Acknowledgments}
\smallskip

One of us (K.P.) wishes to express his gratitude to French Ministry of 
National Education for the attribution of a PAST guest-professor fellowship 
that supported substantially the present collaboration. The authors also 
gratefully acknowledge the financial support by the IN2P3--Polish Laboratories 
Convention as well as the hospitality extended to them during several visits 
at each others laboratories in Lublin and Strasbourg. This work has been 
partly supported by the Polish Committee of Scientific Research under the 
contract No. 2P03B 11519. 
%

%

\end{document}